\documentclass[draft]{agujournal2019}
\usepackage{url} 
\usepackage{lineno}
\usepackage[inline]{trackchanges} 
\usepackage{soul}
\draftfalse

\journalname{Geophysical Research Letters}

\usepackage{amsmath}
\usepackage{amssymb}
\usepackage{chngcntr}
\DeclareSymbolFont{matha}{OML}{txmi}{m}{it}
\DeclareMathSymbol{\varv}{\mathord}{matha}{118}
\DeclareMathSymbol{\varw}{\mathord}{matha}{119}
\newcommand{\mus}{\mu^{*}}
\newcommand{\cs}{c_\mathrm{s}}
\newcommand{\taup}{\tau_\mathrm{p}}
\newcommand{\taur}{\tau_\mathrm{r}}
\newcommand{\tauinf}{\tau_\infty}
\newcommand{\sigmaw}{\sigma_\mathrm{w}}
\newcommand{\Lc}{L_\mathrm{c}}
\usepackage{color}
\newcommand{\modif}[1]{{\color{black}#1}}

\usepackage{moreverb}

\begin{document}

%
%


\title{Earthquake Nucleation along Faults with Heterogeneous Weakening Rate}

%
%





\authors{Mathias Lebihain\affil{1,3}, Thibault Roch\affil{2}, Marie Violay\affil{1}, Jean-Fran\c{c}ois Molinari\affil{2}}

\affiliation{1}{Laboratory of Experimental Rock Mechanics, Civil Engineering Institute, {\'E}cole Polytechnique F{\'e}d{\'e}rale de Lausanne, Station 18, CH-1015 Lausanne, Switzerland}
\affiliation{2}{Computational Solid Mechanics Laboratory, Civil Engineering Institute, Materials Science and Engineering Institute, {\'E}cole Polytechnique F{\'e}d{\'e}rale de Lausanne, Station 18, CH-1015 Lausanne, Switzerland}
\affiliation{3}{Laboratoire Navier, {\'E}cole des Ponts ParisTech, Universit{\'e} Gustave Eiffel, CNRS (UMR 8205), 6-8 avenue Blaise Pascal, 77455 Marne-la-Vall{\'e}e, France}





\correspondingauthor{Mathias Lebihain}{mathias.lebihain@epfl.ch}




\begin{keypoints}
\item The nucleation length of heterogeneous faults with multi-scale asperities of weakening rate can be predicted for slip-dependent friction.
\item Our theory accounts for the transition in fault stability regimes, from the traditional "weakest defect" theory to a homogenized behavior.
\item Only asperities larger than the nucleation length participate actively in the fault stability, while the influence of small heterogeneities can be averaged.
\end{keypoints}

%
%

%
%


\begin{abstract}
The transition from quasi-static slip growth to dynamic rupture propagation constitutes one possible scenario to describe earthquake nucleation. If this transition is rather well understood for homogeneous faults, how the friction properties of \emph{multiscale asperities} may influence the \emph{overall stability} of seismogenic faults remains largely unclear. Combining classical nucleation theory and concepts borrowed from condensed matter physics, we propose a comprehensive analytical framework that predicts the influence of heterogeneities of weakening rate on the nucleation length $\Lc$ for linearly slip-dependent friction laws.  Model predictions are compared to nucleation lengths measured from 2D dynamic simulations of earthquake nucleation along heterogeneous faults. Our results show that the interplay between frictional properties and the asperity size gives birth to three instability regimes (local, extremal, and homogenized), each related to different nucleation scenarios, and that the influence of heterogeneities at a scale far lower than the nucleation length can be averaged.
\end{abstract}

\section*{Plain Language Summary}
Earthquakes occurs on fault. Faults are usually at rest, but they sometimes break and suddenly release a portion of the accumulated elastic energy during \emph{earthquake} ruptures, through radiated waves which may harm populations and structures. Yet, the birth of earthquake (nucleation phase) is not an instantaneous process, and may start with slow slip on the fault. Understanding precisely how this initial phase occurs is thus crucial in predicting earthquake motion. If geophysical models describe it well in an ideal case where the fault is made of the same exact material (homogeneous fault), the role of asperities, which are present from the millimetric rock grain scale to the kilometric tectonic plate scale, remains largely unclear. Here, we propose an extension of the nucleation theory to account for the role of each asperity scale in nucleating earthquakes. Our results show that an earthquake can be triggered by some seismogenic asperities as \emph{previously assumed}, but that these ``weak'' asperities may not control its nucleation if they are small enough. In that case, we show that the birth of earthquakes along complex faults can be accurately studied within the traditional homogeneous nucleation theory.

%
%

\section{Introduction}

Understanding how interfaces fail is of utmost importance in fields ranging from earthquake physics to engineering fracture mechanics. For unstable frictional interfaces such as seismogenic faults, field observations \cite{kato_propagation_2012, bouchon_long_2013} as well as laboratory experiments \cite{dieterich_preseismic_1978, ohnaka_characteristic_1990, ben-david_static_2011, latour_characterization_2013, mclaskey_earthquake_2019} suggest one possible scenario where the onset of fault motion is characterized by the transition from quasi-static slip growth to dynamic rupture propagation \cite{passelegue_dynamic_2016, svetlizky_properties_2016}. The transition happens when a region of critical size $\Lc$ of the fault is slipping. The knowledge of this \emph{nucleation length} proves crucial since it allows to predict both the \emph{loading levels} and the \emph{position} at which earthquake motion starts \cite{uenishi_universal_2003, ampuero_properties_2006, albertini_stochastic_2020}.

Previous theoretical works linked $\Lc$ to the frictional properties of the fault for linear slip-dependent \cite{campillo_initiation_1997, dascalu_fault_2000, uenishi_universal_2003} and more complex rate-and-state \cite{ruina_slip_1983, rubin_earthquake_2005, viesca_stable_2016, aldam_critical_2017, brener_unstable_2018} friction laws along \emph{homogeneous} faults. Yet, 
frictional properties are expected to vary significantly along the fault plane, and with depth due to changes in the local host rock lithology, roughness, or in-situ conditions (normal stress, temperature, pore fluid pressure, etc.) \cite{ohnaka_constitutive_2003, tse_crustal_1986}. Then, how do  these \emph{multi-scale heterogeneous frictional asperities} influence the \emph{global stability} of seismogenic faults? Recent studies \cite{perfettini_scaling_2003, ray_earthquake_2017, dublanchet_dynamics_2018, ray_homogenization_2019, de_geus_how_2019, albertini_stochastic_2020} provide valuable insights on how heterogeneities impact the overall stability of frictional interfaces, but arguably oversimplify the complexity of natural faults by assuming either a homogeneous weakening rate \cite{albertini_stochastic_2020} or orderly placed asperities of uniform size \cite{perfettini_scaling_2003, ray_earthquake_2017, dublanchet_dynamics_2018, ray_homogenization_2019}. A comprehensive framework, which links the variations of frictional properties at all scales to the overall fault stability, is thus dearly lacking.

In this Letter, we build on the theory of static friction \cite{uenishi_universal_2003, rubin_earthquake_2005, viesca_self-similar_2016} and the physics of depinning \cite{tanguy_weak_2004, demery_effect_2014, cao_localization_2018} to develop a theoretical framework that predicts, for \emph{any} heterogeneous linearly slip-dependent fault interface, the critical size $\Lc$ of the earthquake nucleus. Supported by numerical full-field dynamic calculations, we show that the nucleation of an earthquake is not always triggered by the \emph{weakest} heterogeneity, but can also emerge from the \emph{collective} depinning of multiple asperities. We highlight that this shift in instability regime stems from the interplay between the characteristic size of the heterogeneity and the length scale set by the distribution of frictional properties. Finally, we show that, in assessing the stability of an interface, one has to account mainly for perturbations whose wavelength exceeds the nucleation length, since the influence of small-scale asperities can be averaged.

\section{Materials and methods}

\subsection{Dynamic simulations of earthquake nucleation along heterogeneous faults}

\begin{figure}
\noindent\includegraphics[width=\textwidth]{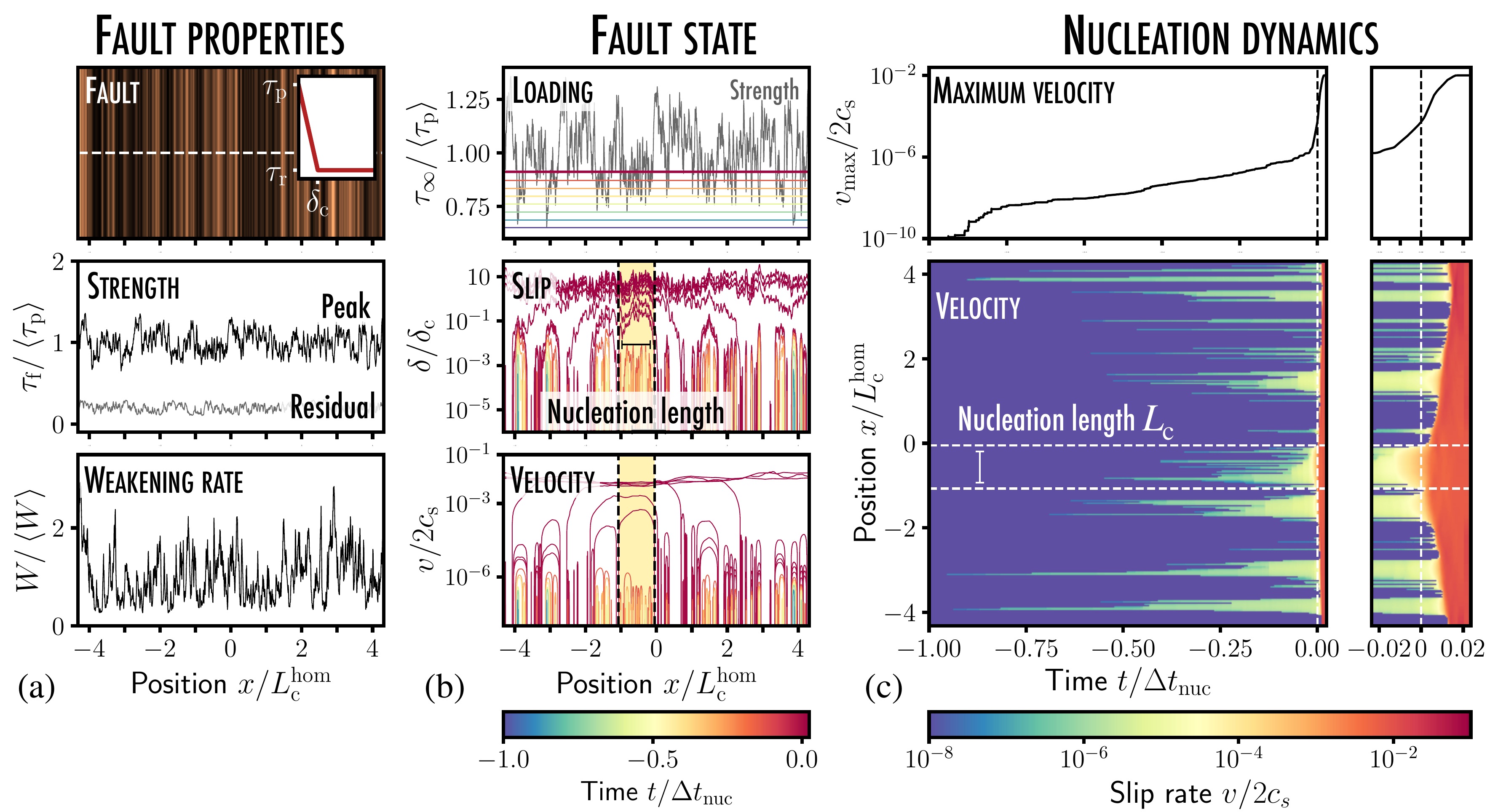}
\caption{(a) Nucleation dynamics along a 1D heterogeneous coplanar fault constituted of brittle (in orange) and ductile (in black) asperities; inset: the fault frictional properties locally follow a linear slip-dependent law; the frictional stress $\tau_\mathrm{f}$ of the interface goes from its peak value $\taup$ to the residual one $\taur$ when the local slip $\delta$ reaches its critical value $\delta_\mathrm{c}$, defining a weakening rate $W= \left(\taup-\taur\right)/\delta_\mathrm{c}$. $\taup$,  $\taur$ and $W$ are varying independently along the fault position $x$. (b) The heterogeneous fault is subjected to a uniform shear loading $\tauinf$. Under the influence of the steadily increasing loading, several regions of the fault start slipping where $\tauinf$ locally exceeds the strength $\taup$. (c) Slip growth develops quasi-statically without any significant velocity burst, until one slip patch reaches a critical length $\Lc$ that leads to the dynamic rupture of the whole interface (see supplemental movie S1).}
\label{fig:HeterogeneousFault}
\end{figure}

We consider two \emph{homogeneous} 2D semi-infinite elastic bodies that are kept in contact with a uniform normal pressure $\sigma_\mathrm{n}$, idealizing the fault structure as a \emph{planar} 1D frictional interface indexed by $x$. The fault is loaded through a macroscopic shear stress $\tauinf\left(x,t\right)$ that slowly increases in time $t$. The friction $\tau_\mathrm{f}$ that opposes interface motion is assumed to be linearly slip-dependent, and fluctuates along the fault (Fig.~\ref{fig:HeterogeneousFault}a). It locally evolves as slip grows from its peak value $\tau_\mathrm{p}\left(x\right)$ to its residual one $\tau_\mathrm{r}\left(x\right)$ with a weakening rate $W(x)$ that describe the material brittleness/ductility. Variations of the frictional properties $\left(\tau_\mathrm{p}, \tau_\mathrm{r}, W\right)$ may emerge along natural faults due to local changes in geometry,  roughness, lithology, or ambient conditions \cite{tse_crustal_1986, ohnaka_constitutive_2003}. Recent works show that nucleation along homogeneous \cite{viesca_self-similar_2016} and heterogeneous \cite{ray_earthquake_2017} faults in the (aging) rate-and-state framework could be investigated from the stability of an \emph{equivalent} interface with \emph{spatially dependent piecewise linear} slip-weakening friction. Despite restrictive assumptions, our work may then provide ways to predict rupture nucleation for more complex and experimentally supported friction laws.

As the macroscopic loading $\tauinf\left(x,t\right)$ grows, it locally exceeds the friction $\tau_\mathrm{p}\left(x\right)$, and the two bodies detach one from another by a slip $\delta\left(x,t\right)$ (Fig.~\ref{fig:HeterogeneousFault}b). Provided that the fault has been at rest for a time far larger than that set by the propagation of elastic waves,  the evolution of $\delta$ is described by the quasi-dynamic equations of elasticity for {Mode II} cracks \cite{rice_spatio-temporal_1993, lapusta_elastodynamic_2000}:
\begin{equation}
\tauinf\left(x,t\right) - \frac{\mus}{2\cs}\frac{\partial \delta}{\partial t}\left(x,t\right) - \mus\mathcal{L}\left[\delta\right]\left(x,t\right) = \max\left[\tau_\mathrm{p}\left(x\right) - W\left(x\right)\delta\left(x,t\right) , \tau_\mathrm{r}\left(x\right) \right]
\label{eq:StressBalance}
\end{equation}
where $\tauinf$ is the far field macroscopic loading, $\cs$ the shear wave velocity, $\mus=\mu/\left(1-\nu\right)$ ($\mu$ and $\nu$ being respectively the shear modulus and the Poisson's ratio), and $\mathcal{L}\left[\delta\right]\left(x,t\right) = \frac{1}{2\pi}\int_{-\infty}^{+\infty} \frac{\partial \delta/\partial x'\left(x',t\right)}{\left(x-x'\right)}dx'$ is a linear operator. In Eq.~\eqref{eq:StressBalance}, the term $-\frac{\mus}{2\cs}\frac{\partial \delta}{\partial t}$, often called ``\emph{radiation damping}'', physically represents wave radiation from the interface to the two elastic bodies, while $\mus\mathcal{L}\left[\delta\right]$ represents the non-local contributions of the overall slip to the local stress state. To investigate the stability of such a heterogeneous fault, we run periodic dynamic simulations building on a spectral boundary integral formulation of fracture \cite{geubelle_spectral_1995, breitenfeld_numerical_1998}. These simulations account for both the static redistribution of stress of Eq.~\ref{eq:StressBalance} and dynamic stress transfers (see supplemental Section S1.2). 

How is the fault stability influenced by the steadily increasing loading? It results in rather complex dynamics as can be observed in Fig.~\ref{fig:HeterogeneousFault}b.  multiple regions slipping at an accelerated rate, referred to as ``\emph{slip patches}'', start to nucleate on the positions where $\tau_\mathrm{p}$ is low. As the loading is further increased, they grow quasi-statically, and coalesce into larger slipping regions. This initial nucleation stage of duration $\Delta t_\mathrm{nuc}$ proves rather quiescent since no major velocity burst is observed. Yet, at $t=0$, an instability develops on the right part the fault: a rupture propagates dynamically, and the two bodies start sliding one onto another at an uniform slip rate.

If such a simulation constitutes one realistic scenario for natural earthquakes nucleation, the simultaneous growth of multiple slip patches prevents any accurate measurement of the size $\Lc$ of the instability nucleus, which might well be twice as large as our measurement of Fig.~\ref{fig:HeterogeneousFault}c. Yet, identifying this critical length scale proves crucial since it gives access to (i) the loading levels \cite{uenishi_universal_2003} and (ii) the position at which an earthquake nucleates \cite{ampuero_properties_2006, albertini_stochastic_2020} when $W$ is homogeneous along the fault.

These difficulties arise from spatial variations of peak strength that have been proven to play no role in the stability behavior of a slip-dependent frictional interface, which is solely controlled by the weakening rate $W$ \cite{favreau_initiation_1999, uenishi_universal_2003}. Indeed, assuming that the macroscopic loading $\tauinf$ increases slowly enough and that the slip perturbation is small enough, the interface velocity $v=\dot{\delta}$ is described near the instability by \cite{uenishi_universal_2003}:
\begin{equation}
\frac{\mus}{2\cs}\frac{\partial v}{\partial t}\left(x,t\right) + \mus\mathcal{L}\left[v\right]\left(x,t\right) - W\left(x\right)v\left(x,t\right) = 0
\label{eq:VelocityBalance}
\end{equation}
where only $W$ is involved. This observation is supported by recent numerical simulations of crack nucleation along interfaces with stochastic distributions of $\tau_\mathrm{p}$ and homogeneous $W$ \cite{albertini_stochastic_2020}, except in rare situations where the asperity scale interacts with the nucleation length \cite{schar_nucleation_2021}. One may then focus on variations of weakening rate $W$ to quantify the influence of multi-scale heterogeneities on fault stability.

\subsection{Measuring the nucleation length in presence of weakening rate variations: a model fault approach}

\begin{figure}
\noindent\includegraphics[width=\textwidth]{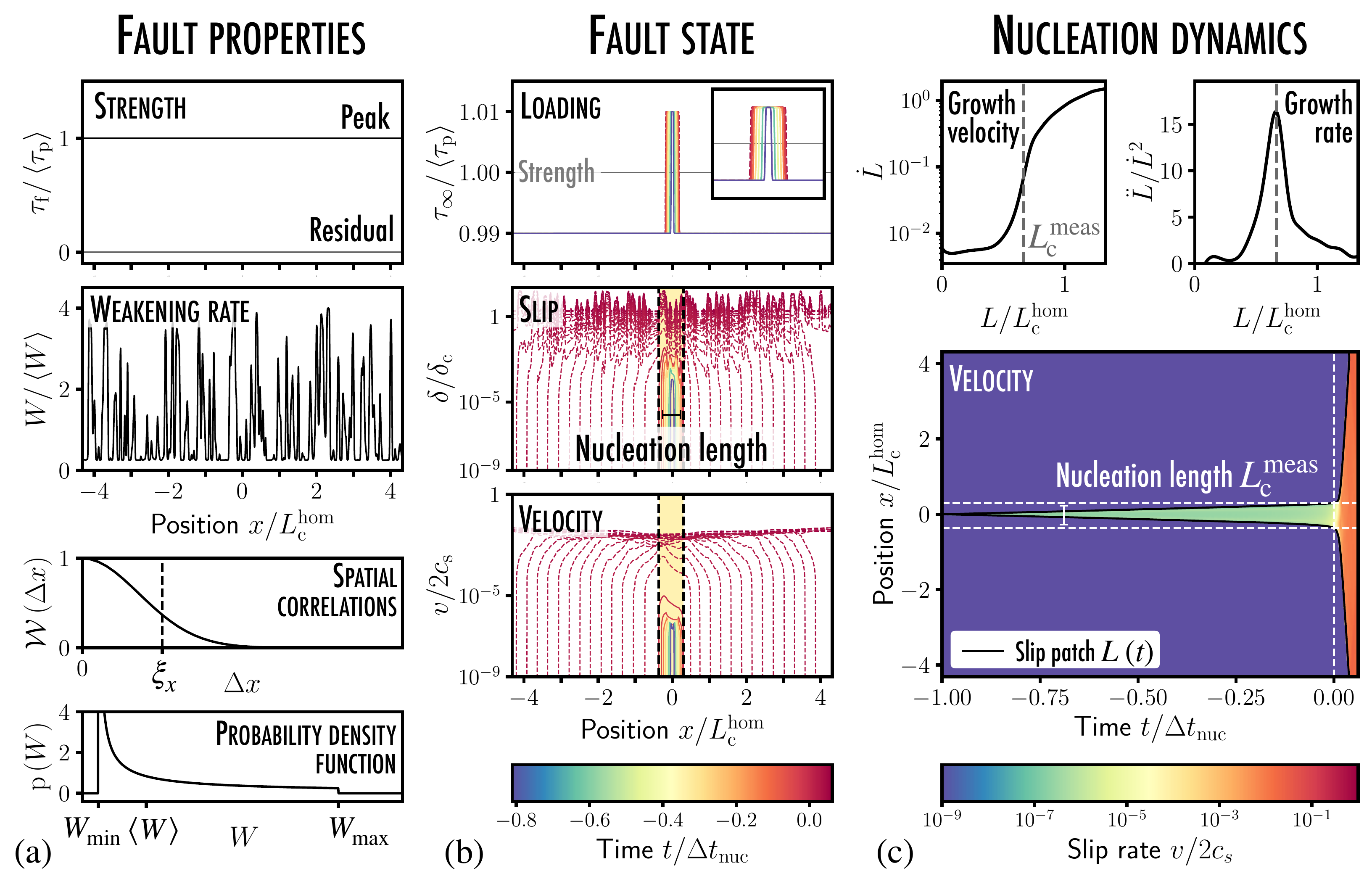}
\caption{Measuring the nucleation length of a heterogeneous fault: (a) $\taup$ and $\taur$ are considered uniform along the interface, while the weakening rate $W$ varies with the position. These variations occur over a characteristic length $\xi_x$, and are distributed following a beta-distribution $\mathrm{p}\left(W\right)$ between two extremal values $\left[W_\mathrm{min},W_\mathrm{max}\right]$. (b) The model interface is loaded through an over-stressed patch that slowly expands in time. A slip perturbation $\delta$ and an associated velocity perturbation $v=\dot{\delta}$ develop as time grows. The dynamics are characterized by two phases: the first phase consists of quasi-static growth (solid lines) and the second involves dynamic crack propagation (dashed lines) when the slipping region reaches a critical size $\Lc$. (c) This shift in dynamics is observed on the temporal evolution of the slip patch size $L(t)$, or its growth velocity $\dot{L}$. In particular, $\dot{L}$ hits an inflection point at $L=\Lc$ (in linear-log space), which provides an accurate measurement of $\Lc$ from the growth rate $\ddot{L}/\dot{L}^2$. See movie~S2, and Figure~S2 and movie S3 for comparison with the homogeneous case of \cite{uenishi_universal_2003}.}
\label{fig:ModelFault}
\end{figure}

We thus focus on the stability behavior of an idealized fault along which both the peak $\tau_\mathrm{p}$ and the residual friction $\tau_\mathrm{r}$ are uniform (Fig.~\ref{fig:ModelFault}a). To make any parallel to Mode I fracture easier, and without any loss of generality, we set $\tau_\mathrm{r}\left(x\right)=0$ \cite{albertini_stochastic_2020}. Meanwhile, the weakening rate $W$ may vary from several orders of magnitude along the fault. Following the procedure of \cite{albertini_stochastic_2020} (see supplemental Section S1.1), we generate $W$ fields that follow Gaussian correlations up to a characteristic length scale $\xi_x$. Moreover, the values of $W$ follow a beta distribution of average $\left<W\right>$ and standard deviation $\sigmaw$, between two extremal values $\left[W_\mathrm{min},W_\mathrm{max}\right]$.  We set the nucleation length of the reference \emph{homogeneous} material with uniform $\left<W\right>$ as the adimensionalizing length of the system $\Lc^\mathrm{hom} \simeq 1.158 \mus/\left<W\right>$ \cite{uenishi_universal_2003}. In the following, we consider $\sigmaw = \left<W\right>$, $W_\mathrm{min} = 0.25\left<W\right>$, $W_\mathrm{max} = 4\left<W\right>$, and $\xi_x = 0.05\Lc^\mathrm{hom}$.  The behavior of such a heterogeneous interface remains out of scope of the current theories of rupture nucleation where $W$ is homogeneous \cite{favreau_initiation_1999, uenishi_universal_2003, ampuero_properties_2006,  albertini_stochastic_2020}. We then wonder how \emph{local} variations of $W$ as well as their intensity impact the \emph{overall} fault stability.

In presence of spatial variations of $W$, the nucleation length is expected to fluctuate along the fault. In order to investigate the local instability dynamics, we force nucleation at a given point, referred to as ``fault center'', by considering a macroscopic loading consisting in a slowly expanding region of size $L_\sigma=c_\sigma t$ ($c_\sigma \ll c_\mathrm{s}$), where the stress locally exceeds the frictional resistance $\tauinf \geq \tau_\mathrm{p}$ (Fig.~\ref{fig:ModelFault}b). We observe in Fig.~\ref{fig:ModelFault}b that a typical nucleation event is very similar, yet much simpler, to that of the heterogeneous fault of Section~2.1. Its dynamics consists of two distinct regimes: (i) the first regime involves \emph{stable quasi-static slip growth} for $t<0$ where a portion $L$ of the interface is slipping , while (ii) the second involves \emph{unstable dynamic crack propagation} for $t>0$ where a rupture front propagates until the whole fault is moving. The shift from one regime to another occurs when the slipping region outgrowths a critical length $\Lc$, independently of the nature of the loading shape as long as it is \emph{peaked} (see Fig.~S4 in supplemental). Importantly, this instability results from the collective motion of multiple asperities ($\Lc\simeq 13 \xi_x$ in Fig.~\ref{fig:ModelFault}). 

To quantify the influence of spatial variations of $W$ on $\Lc$, we first propose a \emph{heuristic} framework to measure it from numerical calculations. Looking at the evolution of the slip patch size $L$ over time in Fig.~\ref{fig:ModelFault}c, we observe that its growth velocity $\dot{L}$ follows an S-shaped curve and hits an inflexion point (in linear-log space) when $L=\Lc$, as previously observed in laboratory experiments of earthquake nucleation between two polycarbonates blocks \cite{latour_characterization_2013}. The nucleation length $\Lc$ may then be estimated from the maximal growth rate $\ddot{L}/\dot{L}^2$, as the length where the patch expansion is at its strongest. The validity of our heuristic approach is assessed on homogeneous interfaces for which $\Lc^\mathrm{hom}\simeq 1.158 \mus/\left<W\right>$  is known \emph{a priori} \cite{favreau_initiation_1999, uenishi_universal_2003}. We use it to estimate numerically the critical length $\Lc$ of heterogeneous interfaces with a $\pm 5\%$ precision, corresponding to the error observed for the homogeneous interface of known $\Lc$ (see supplemental Section S2.1).

\section{Results and discussion}

\subsection{Theoretical model for nucleation length predictions}

Here, we propose a way to determine the critical length $\Lc$ \emph{analytically} building on both the theory of static friction and the physics of depinning. For a velocity perturbation $v$ centered in $x=x_0$ with a support of size $L$, Eq.~\eqref{eq:VelocityBalance}  becomes:
\begin{equation}
\frac{\mus}{2\cs}\frac{\partial v}{\partial t}\left(X,t\right) + \frac{2\mus}{L}\mathcal{L}_{1}\left[v\right]\left(X,t\right) - W\left(x_0+LX/2\right)v\left(X,t\right) = 0
\label{eq:LSABalance}
\end{equation}
where $\mathcal{L}_1\left[v\right]\left(x,t\right) = \frac{1}{2\pi}\int_{-1}^{+1} \frac{\partial v/\partial X'\left(X',t\right)}{\left(X-X'\right)}dX'$ is the linear operator introduced in \cite{dascalu_fault_2000, uenishi_universal_2003}, and $X=2\left(x-x_0\right)/L$ is the reduced position. 

To assess the fault stability, we perform a Linear Stability Analysis on Eq.~\eqref{eq:LSABalance}. It consists of finding the perturbation size $L$ for which the linear symmetric operator $\mathcal{D}[v]\left(X\right) = \frac{2\mus}{L}\mathcal{L}_{1}\left[v\right]\left(X\right)-W\left(LX/2\right)v\left(X\right)$ admits a zero eigenvalue. Condensed matter physics provides one way to tackle this problem in presence of heterogeneities \cite{tanguy_weak_2004, demery_effect_2014, cao_localization_2018}: we expand the perturbation $v$ with the disorder intensity $\sigmaw$ up to 2\textsuperscript{nd} order $v=v_0+\sigmaw v_1+\sigmaw^2 v_2$, and solve the eigenproblem $\mathcal{D}\left[v\right]=\omega v$, where $\omega=\omega_0+\sigmaw \omega_1+\sigmaw^2 \omega_2$ (see supplemental Section S3). The nucleation length $\Lc$ is the solution of the transcendental equation~\eqref{eq:NucleationLength}, which encompasses the main novelty of the paper.

\begin{flalign}
 \label{eq:NucleationLength}
\frac{2\lambda_0\mus}{\Lc\left(x_0\right)} & - \int_{-1}^{+1} W\left(x_0+\Lc\left(x_0\right) X'/2\right) \nu_0\left(X'\right)^2 dX' &\\\nonumber
& - \frac{\Lc\left(x_0\right)}{2\mus} \sum\limits_{1 \leq k\leq k_\mathrm{c}}\frac{1}{\lambda_k-\lambda_0} \left[\int_{-1}^{+1} W\left(x_0 + \Lc\left(x_0\right) X'/2\right) \nu_0\left(X'\right)\nu_k\left(X'\right) dX'\right]^2  = 0 &
\end{flalign}

In Eq.~\eqref{eq:NucleationLength}, $\nu_k$ denotes the  $k^\mathrm{th}$ eigenmode associated to the eigenvalue $\lambda_k$ of the homogeneous eigenproblem $\mathcal{L}_1\left[v\right]=\lambda_k v$ \cite{dascalu_fault_2000, uenishi_universal_2003}. The first two terms of Eq.~\eqref{eq:NucleationLength} represent the heterogeneities contributions up to the first order. The value of the critical length $\Lc$ at a position $x=x_0$ involves spatial variations of $W$ on scale potentially larger than the heterogeneity size $\xi_x$. This collective yet heterogeneous behavior in earthquake nucleation cannot be grasped by the homogeneous $W$ theory \cite{uenishi_universal_2003, albertini_stochastic_2020}. The third term corresponds to second-order contributions up to a critical mode $k_\mathrm{c} \simeq 2\Lc/\xi_x$. This higher order term accounts for the influence of the spatial shape of $W$ in all its complexity, beyond the special cases of periodic ordered distributions of asperities \cite{perfettini_scaling_2003, ray_earthquake_2017, dublanchet_dynamics_2018, ray_homogenization_2019}. Eq.~\eqref{eq:NucleationLength}, is only valid as long as no point of the fault reaches its residual friction value (i.e. $\delta(x,t) < \delta_\mathrm{c}(x)$). But $W(x$ can be replaced by the instantaneous weakening rate $W\left[x,\delta(x,t)\right]$ to assess fault stability around a stable slip state $\delta(x,t)$ in the case of \emph{non-linear} or \emph{piecewise linear} slip-dependent friction. Yet, if Eq.~\eqref{eq:NucleationLength} gives \emph{qualitative} information on the influence of a non-stationary weakening onto the nucleation process, it does not provide a \emph{quantitative} framework to predict rupture nucleation for these more complex friction laws, as the slip evolution remains unknown.

When we compare the theoretical predictions $\Lc^\mathrm{pred}$ of Eq.~\eqref{eq:NucleationLength} to numerically estimated critical lengths $\Lc^\mathrm{meas}$ for an asperity size $\xi_x$ that varies over 4 orders of magnitude, we observe an excellent agreement (Fig.~\ref{fig:InstabilityRegimes}a). Note that (i) the nucleation length cannot be estimated from \cite{uenishi_universal_2003}'s homogeneous theory (Fig.~\ref{fig:InstabilityRegimes}b) and (ii) second-order contributions are required for accurate predictions (see Fig.~S7 in supplemental).

Equation~\eqref{eq:NucleationLength} unveils rich physics about the impact of microscopic heterogeneities on the macroscopic fault stability. From it, one can directly link spatial profiles of weakening rate $W$ to the local evolution of the nucleation length $\Lc$ along the interface (see black solid lines in Fig.~\ref{fig:InstabilityRegimes}c-e). In our simulations, the effective nucleation length corresponds to the one predicted at $x=0$ due to the \emph{peaked} nature of the loading. In more realistic cases,  the position $x_0$ of the earthquake nucleus (and the associated nucleation length $\Lc\left(x_0\right)$) will depend on (i) heterogeneities of peak strength $\taup$,  but also (ii) on the spatial shape of the macroscopic loading $\tau_\infty$, similarly to what has been observed for non-linear slip-weakening laws \cite{rice_rupture_2010}.

Overall, our framework provides ways to quantify the influence of a single heterogeneity on the fault stability depending on its size and intensity (see Fig.~\ref{fig:InstabilityRegimes}), as well as that of the superposition of multiple perturbations of frictional properties (see Fig.~\ref{fig:ModalSuperposition}). Next, we build on Eq.~\eqref{eq:NucleationLength} and distinguish three instability regimes that can be linked to realistic earthquake nucleation scenarios on natural faults.

\subsection{Instability regimes in earthquake nucleation}

\begin{figure}
\noindent\includegraphics[width=\textwidth]{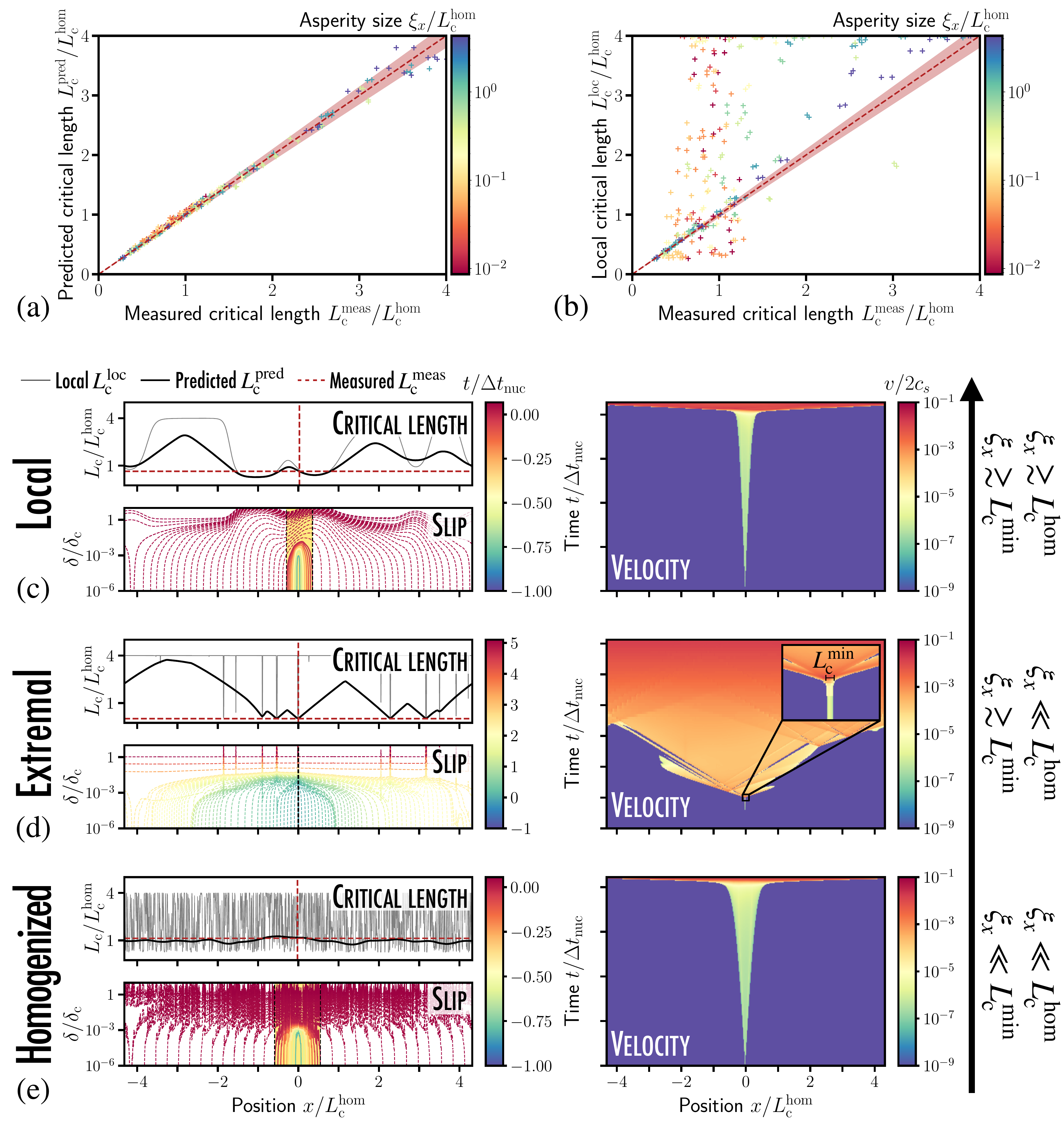}
\caption{(a) The length of the critical instability nucleus $\Lc^\mathrm{meas}$ measured from the dynamic simulations is compared to the theoretical prediction $\Lc^\mathrm{pred}$ of Eq.~\eqref{eq:NucleationLength} at the fault center $x=0$, for a broad range of characteristic scale $\xi_x$ of the asperities (320 simulations). Red region: $\pm 5\%$ error on $\Lc^\mathrm{meas}$. (b) $\Lc^\mathrm{meas}$ may strongly differ from the local nucleation length of \cite{uenishi_universal_2003} $\Lc^\mathrm{loc}(x)\simeq 1.158\mus/W(x)$ at $x=0$. The interplay between the length scales set by the frictional properties and the asperity size gives birth to three instability regimes: (c) when $\xi_x$ is larger than the homogeneous nucleation length $\Lc^\mathrm{hom}\simeq 1.158\mus/\left<W\right>$ set by the average frictional properties, the effective nucleation length $\Lc\left(x\right)$ of  Eq.~\eqref{eq:NucleationLength} follows \cite{uenishi_universal_2003}'s predictions $\Lc\left(0\right) \simeq \Lc^\mathrm{loc}\left(0\right)$; (d) when $\xi_x$ is smaller than $\Lc^\mathrm{hom}$ yet larger than the minimal nucleation length $\Lc^\mathrm{min}\simeq 1.158\mus/W_\mathrm{max}$ set by the most brittle defect, $\Lc\left(x\right)$ departs significantly from $\Lc^\mathrm{loc}\left(x\right)$ but can be locally controlled by the \emph{extrema} of the weakening rate distribution $\Lc\left(0\right) \simeq \Lc^\mathrm{min}$. Inset: instability birth at $L\simeq \Lc^\mathrm{min} = 0.02 \Lc^\mathrm{hom}$; (e) when $\xi_x$ is smaller than both $\Lc^\mathrm{hom}$ and $\Lc^\mathrm{min}$, the nucleation behavior is \emph{homogenized} and the nucleation length $\Lc\left(x\right)$ is comparable to that set by the average frictional properties $\Lc\left(x\right) \simeq \Lc^\mathrm{hom}$ (see supplemental movies S4/S5/S6).}
\label{fig:InstabilityRegimes}
\end{figure}

In natural fault zones, heterogeneities in friction occur over many different scales. We observe them at the \emph{centimetric} scale with minerals, clasts and foliation, at the \emph{metric/decametric} scale along large faults consisting of different lithologies, up to the scale of tectonic plates where \emph{kilometric} asperities generated by heterogeneous stress distribution have been suggested as potential nucleation sites for megathrust earthquakes in subduction zones. It is still uncertain how those different scales may interact with each other, and how they ultimately impact the nucleation of earthquakes. Building on Eq.~\eqref{eq:NucleationLength}, we highlight in Fig.~\ref{fig:InstabilityRegimes}c-e three different instability regimes, referred to as local, extremal and homogenized regimes. They emerge from the interplay between three length scales: the heterogeneity size $\xi_x$, the nucleation length associated to average frictional properties $\Lc^\mathrm{hom}\simeq 1.158\mus/\left<W\right>$, and the scale set by the weakest defect along the fault $\Lc^\mathrm{min} \simeq 1.158\mus/W_\mathrm{max}$. 

(i) Local regime: when $\xi_x  \gtrsim \Lc^\mathrm{hom}$ and $\xi \gtrsim \Lc^\mathrm{min}$, the weakening rate $W$ is almost constant over the nucleation patch (see Fig.~\ref{fig:InstabilityRegimes}c). One then retrieves the dynamics of homogeneous nucleation \cite{uenishi_universal_2003}, and the effective nucleation length is set by the \emph{local} frictional properties at the fault center $\Lc\left(0\right) \simeq \Lc^\mathrm{loc} \simeq 1.158\mus/W\left(0\right)$, which can be distributed above ($W\left(0\right) < \left<W\right>$) or below ($W\left(0\right) > \left<W\right>$)  $\Lc^\mathrm{hom}$.

(ii) Extremal regime: when $\xi_x  \ll \Lc^\mathrm{hom}$ and $\xi_x \gtrsim \Lc^\mathrm{min}$, \modif{a critical nucleation patch of size $\Lc\left(0\right) \simeq \Lc^\mathrm{min}$ may develop within a single brittle asperity of size $\xi_x$}, where the weakening rate reaches its \emph{maximal} value $W_\mathrm{max}$. This small event destabilizes the interface as a whole, generating a complex dynamics of multiple slip pulses (see velocity map in Fig.~\ref{fig:InstabilityRegimes}d \modif{for which $\xi_x = 0.01 \Lc^\mathrm{hom} = 2 \Lc^\mathrm{min}$)}. Along natural faults, these small ruptures \modif{may} be arrested by local barriers of strength $\taup$, but they may trigger a \emph{cascade} of nucleation events centered on other weakest spots until the entire fault fails \cite{zhang_heterogeneous_2003, noda_large_2013, de_geus_how_2019}. Note that (1) these brittle asperities influence the effective nucleation length $\Lc$ far away from them, and that, in contrast, and (2) other weak spots may not be critical if not brittle enough i.e. $\xi_x \ll \Lc^\mathrm{loc} \gtrsim \Lc^\mathrm{hom}$ (see \modif{the spatial evolution of $\Lc^\mathrm{pred}$ in black line} in Fig.~\ref{fig:InstabilityRegimes}d and supplemental Section S4).

(iii) Homogenized regime:  when $\xi_x  \ll \Lc^\mathrm{hom}$ and $\xi_x \ll \Lc^\mathrm{min}$, no critical slip patch can develop within a single asperity. Nucleation occurs after the collective depinning of multiple asperities \cite{perfettini_scaling_2003, dublanchet_dynamics_2018, ray_homogenization_2019} with dynamics similar to that of homogeneous nucleation. The critical length $\Lc$ fluctuates around its \emph{homogenized} value $\Lc^\mathrm{hom}$ set by the averaged frictional properties (see Fig.~\ref{fig:InstabilityRegimes}e), and can then be studied within the homogeneous nucleation theory of \cite{uenishi_universal_2003}. Note that Eq.~\ref{eq:NucleationLength} fully capture fluctuations of $\Lc$ around $\Lc^\mathrm{hom}$, which may grow as the intensity $\sigmaw$ of weakening rate fluctuations increases.

We argue here that all three instability regimes could occur along natural faults depending on their size, geometry, maturity, and lithology. But the \emph{homogenized} regime proves to be of major importance for geophysical applications. Indeed, heterogeneous fracture is often described as a \emph{critical} phenomenon controlled by the \emph{weakest} defect, thus ruling out its study within a homogeneous framework. Yet, our results suggest that under the scale separation condition $\xi_x \ll \Lc^\mathrm{min}$, the stability behavior of a heterogeneous fault can be studied within the homogeneous framework of \cite{favreau_initiation_1999} and \cite{uenishi_universal_2003} with $\Lc \simeq \Lc^\mathrm{hom} \simeq 1.158\mus/\left<W\right>$. Moreover, the existence of the \emph{homogenized} regime may account for the relative reproductivity of laboratory experiments where sample roughness is often imposed and kept relatively smooth, and further justifies their relevance in the modeling of natural faults.

\subsection{Influence of each asperity scale to the global stability of heterogeneous fault}
So far we considered cases where the distribution of weakening rate asperities could be described through a unique length scale $\xi_x$. Yet, heterogeneities of weakening rate may emerge from e.g. the fault roughness that exhibits a scale-free self-affine behavior that spans over several decades of length scales \cite{candela_roughness_2012}, which makes the modeling of rough faults particularly challenging from a numerical point of view. Up to now,  it is still largely unclear which length scales actively participate in the fault stability and which may be averaged in a realistic modeling of earthquake nucleation along rough faults.

\begin{figure}
\noindent\includegraphics[width=\textwidth]{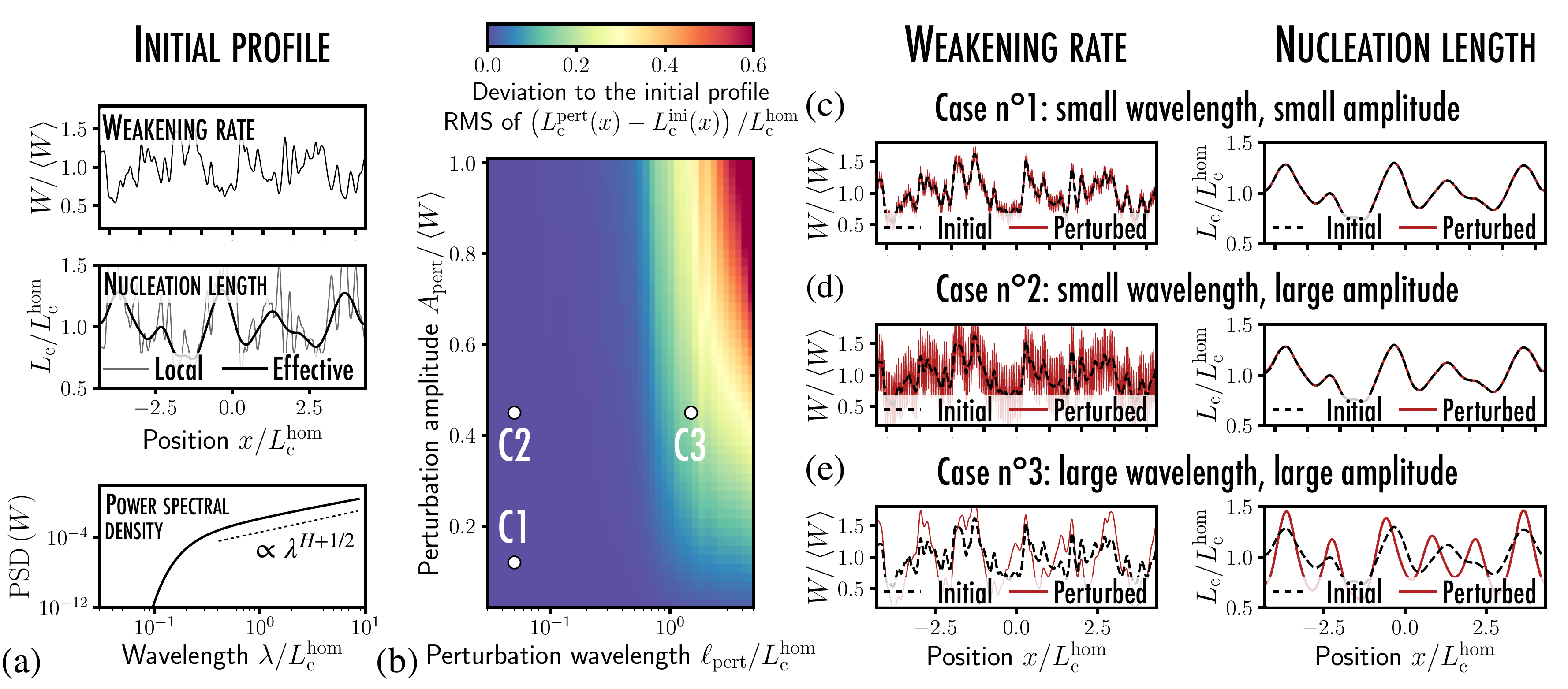}
\caption{Influence of a modal perturbation on the stability of a heterogeneous fault: (a) an initial weakening rate profile $W_\mathrm{ini}\left(x\right)$ consisting of the superposition of multiple spatial modes is considered. The spatial variations of $W_\mathrm{ini}$ gives birth to an effective profile of nucleation length $\Lc^\mathrm{ini}\left(x\right)$ following Eq.~\eqref{eq:NucleationLength}. (b) The influence of the superposition of a unimodal perturbation $\hat{w}$ of period $\ell_\mathrm{pert}$ and amplitude $A_\mathrm{pert}$ to the initial $W$-profile is quantified through the \modif{root-mean-square of the difference between} the spatial profile of initial nucleation length $\Lc^\mathrm{ini}\left(x\right)$ related to $W_\mathrm{ini}\left(x\right)$, and the perturbed one $\Lc^\mathrm{pert}\left(x\right)$ related to $W_\mathrm{ini}\left(x\right) + \hat{w}\left(x\right)$. (c-e) When the perturbation wavelength $\ell_\mathrm{pert}$ is smaller than the initial nucleation length $\Lc^\mathrm{ini}$, it does not change its spatial profile, no matter the perturbation amplitude. Only larger wavelength perturbations may influence the nucleation length.}
\label{fig:ModalSuperposition}
\end{figure}

To further demonstrate the potential of our theoretical framework, we consider a heterogeneous fault with a weakening rate profile $W_\mathrm{ini}\left(x\right)$ (see Fig.~\ref{fig:ModalSuperposition}a), that emerges from a multi-scale distribution of  asperities with a Hurst exponent $H=0.7$ \cite{ampuero_properties_2006}. The nucleation length $\Lc^\mathrm{ini}\left(x\right)$ can be computed from Eq.~\eqref{eq:NucleationLength}. We superpose to the initial weakening rate profile $W_\mathrm{ini}$ a unimodal perturbation $\hat{w}$ of period $\ell_\mathrm{pert}$ and amplitude $A_\mathrm{pert}$, giving birth to a perturbed profile $W_\mathrm{pert}$:
\begin{equation}
W_\mathrm{pert}\left(x\right) = W_\mathrm{ini}\left(x\right)+A_\mathrm{pert} \cdot \mathrm{cos}\left(\dfrac{2\pi}{\ell_\mathrm{pert}}x\right)
\end{equation}

When computing the nucleation length $\Lc^\mathrm{pert}\left(x\right)$ associated to $W_\mathrm{pert}\left(x\right)$, we observe that only perturbations whose wavelength is larger than the reference nucleation length $\Lc^\mathrm{ini}\left(x\right)$ matter (see Fig.~\ref{fig:ModalSuperposition}b\& e), and that the perturbation of critical length increases with both the wavelength $\ell_\mathrm{pert}$ and the amplitude $A_\mathrm{pert}$. Furthermore, small-scale perturbations ($\ell_\mathrm{pert} \lesssim \Lc^\mathrm{hom}$) do not change the nucleation length $\Lc^\mathrm{pert}$, whether its amplitude is small (Fig.~\ref{fig:ModalSuperposition}c) or large (Fig.~\ref{fig:ModalSuperposition}d). Note that, if the initial nucleation length $\Lc^\mathrm{ini}$ locally drops to extremal values (see Fig.~\ref{fig:InstabilityRegimes}d), very small-scale asperities may then influence the overall stability behavior of the fault. 

Overall, our work provides then \emph{quantitative reasoning} to assess which scale of asperities should be included in the modeling of complex faults and which can be \emph{averaged}, when frictional heterogeneities span over several length scales.

\section{Conclusion}

Nucleation processes along fault with differential frictional weakening $W$ is a \emph{collective} phenomenon that may involve the progressive depinning of multiple asperities until a perturbation of size $\Lc$ is reached. Building on the theory of static friction and the physics of depinning, we proposed an analytical framework that allows to predict the effective critical length $\Lc$ for \emph{any} spatial profile of $W$. This framework has been successfully compared to dynamic simulations of Mode II friction, and is directly tractable to nucleation of Mode I and Mode III fracture along weak interfaces. It provides clues to explain various nucleation scenarios observed in laboratory experiments and in nature, as well as to derive scale-separation conditions assessing the influence of one asperity scale on the overall fault stability. Further, \modif{it may provide ways to estimate} the \emph{shear loading levels} and the \emph{position} at which nucleation occurs along more complex interfaces \emph{of known frictional properties}, where all frictional quantities $\left(\taup, \taur, W\right)$ as well as the external loading $\left(\sigma_\mathrm{n}, \tau_\infty\right)$ fluctuate due to e.g. fault roughness \cite{cattania_precursory_2021}. The recent analogy drawn between earthquake nucleation for rate-and-state friction and rupture initiation along heterogeneous piecewise linear slip-weakening interfaces \cite{viesca_self-similar_2016, ray_earthquake_2017} provides convincing ways to extend the proposed framework to rate-and-state friction laws. The generalization of our results to (1) non-linear slip-dependent friction laws, and (2) a three-dimensional setting is not straightforward, but one may adapt the approach proposed in this work to handle fault inhomogeneities to the energetic nucleation framework of \cite{rice_rupture_2010}. \modif{Yet, further experimental work is needed to assess the validity of our framework in predicting the influence of heterogeneities on the nucleation process. Dynamic rupture experiments performed on model micro-architectured faults in the laboratory (like e.g. \cite{berman_dynamics_2020}) may constitute a critical test for our theory.}

\acknowledgments
Data regarding the figures of the main text are available on Zenodo \cite{lebihain_data_2021}. M.L. acknowledges funding provided by the Swiss National Science Foundation (Grant  CRSK-2\_190805). M.V. and M.L. acknowledge support provided by the ERC BEFINE (Grant 757290). The authors thank Dr. F.X. Passel{\`e}gue for fruitful discussions and a critical reading of the manuscript, as well as the three anonymous reviewers whose valuable comments improved the manuscript. This work benefits from discussions with Dr. G. Albertini and Dr. D.S. Kammer.


\end{document}